\documentclass{siamonline0516}  
\RequirePackage{graphicx}
\RequirePackage{latexsym}
\usepackage[T1]{fontenc}
\usepackage[utf8]{inputenc}
\usepackage[english]{babel}
\usepackage{amsmath,amssymb}
\usepackage{color}
\usepackage{algorithm,algorithmicx,algpseudocode}
\usepackage{setspace}
\newtheorem{thm}{Theorem}

\newtheorem{remark}[thm]{Remark}

\newcommand{\rmd}{\mathrm{d}}
\newcommand{\rme}{\mathrm{e}}

\newcommand{\rmq}{\mathrm{q}}
\newcommand{\bfb}{\mathbf{b}}

\newcommand{\bfp}{\mathbf{p}}
\newcommand{\bfq}{\mathbf{q}}

\newcommand{\bfu}{\mathbf{u}}

\newcommand{\bfx}{\mathbf{x}}

\newcommand{\bfz}{\mathbf{z}}
\newcommand{\bfzero}{\mathbf{0}}
\newcommand{\bfA}{\mathbf{A}}
\newcommand{\bfB}{\mathbf{B}}
\newcommand{\bfC}{\mathbf{C}}
\newcommand{\bfD}{\mathbf{D}}
\newcommand{\bfP}{\mathbf{P}}
\newcommand{\bfQ}{\mathbf{Q}}
\newcommand{\bfR}{\mathbf{R}}

\newcommand{\bfW}{\mathbf{W}}
\newcommand{\bfZ}{\mathbf{Z}}

\newcommand{\calD}{\mathcal{D}}
\newcommand{\calE}{\mathcal{E}}
\newcommand{\calF}{\mathcal{F}}

\newcommand{\calL}{\mathcal{L}}
\newcommand{\calO}{\mathcal{O}}
\newcommand{\calR}{\mathcal{R}}

\newcommand{\calT}{\mathcal{T}}
\newcommand{\bbE}{\mathbb{E}}
\newcommand{\bbP}{\mathbb{P}}
\newcommand{\vecr}{\vec{r}}
\newcommand{\Var}{\mathrm{Var}}
\newcommand{\He}{\mathit{He}}

\newcommand{\T}{^\mathsf{T}}
\newcommand{\iT}{^{-\mathsf{T}}}
\newcommand{\Mh}{M_\mathrm{h}}

\newcommand{\ds}{\displaystyle}
\newcommand{\topic}[1]{}

\newcommand{\todo}[1]%
{\textcolor{red}{\fbox{to do}}#1\textcolor{red}{\fbox{end}}}

\begin{document}
	
	\title{Choice of Damping Coefficient in Langevin Dynamics}
	%\titlerunning{Choice of Damping Coefficient in Langevin Dynamics}
	\author{Robert D. Skeel\thanks{School of Mathematics and Statistical Sciences, Arizona State University, 900 S Palm Walk, Tempe, AZ 85281, USA, E-mail: \email{rskeel@asu.edu}} \and Carsten Hartmann\thanks{Institute of Mathematics, Brandenburgische Technische Universit\"at Cottbus-Senftenberg, 03046 Cottbus, Germany, E-mail: \email{hartmanc@b-tu.de}}}

	\date{\today}
	
	\maketitle
	
	\begin{abstract}
		This article considers the application of Langevin dynamics to sampling
		and investigates how to choose the damping parameter
		in Langevin dynamics for the purpose of maximizing thoroughness
		of sampling. Also, it considers the computation of measures of
		sampling thoroughness.
	\end{abstract}

	\section{Introduction}
	
	\topic{importance}
	Langevin dynamics is a popular tool for molecular simulation.
	It requires the choice of a damping coefficient,
	which is the reciprocal of a diffusion coefficient.
	(More generally this might be a diffusion tensor.)
	The special case of a constant scalar diffusion coefficient
	is the topic of this article.
	The motivation for this study is a suspicion
	that proposed novel MCMC propagators based on Langevin dynamics
	(in particular, stochastic gradient methods
	for machine learning~\cite{HighLow2020,DFBC14})
	might be obtaining their advantage
	at the expense of reduced sampling efficiency,
	as, say, measured by effective sample size.
	
	\topic{dynamics}
	For simulations intended to model the dynamics,
	the appropriate choice of $\gamma$ is based on physics.
	Generally, the dissipation and fluctuation terms
	are there to account for omitted degrees of freedom.
	In their common usage as thermostats, they
	model the effect of forces due to atoms just outside the set
	of explicitly represented atoms.
	These are essentially boundary effects,
	which disappear in the thermodynamic limit $N_\mathrm{atoms}\rightarrow\infty$,
	where $N_\mathrm{atoms}$ is the number of explicitly represented atoms.
	Since the ratio of the number of boundary atoms to interior atoms
	is of order $N_\mathrm{atoms}^{-1/3}$,
	it might be expected
	that $\gamma$ is chosen to be proportional to $N_\mathrm{atoms}^{-1/3}$.
	%(It is interesting to note that constrained dynamics and united atom
	%force fields are also instances, though less severe,
	%of omitted degrees of freedom.)
	There is second possible role for the addition
	of fluctuation-dissipation terms in a dynamics simulation:
	with a small damping coefficient,
	these terms can also play a role in stabilizing
	a numerical integrator \cite{ICWS01},
	which might be justified if the added terms
	are small enough to have an effect no greater
	than that of the discretization error.
	
	\topic{sampling}
	The bulk of molecular simulations, however, are ``simply'' for the purpose
	of drawing random samples from a prescribed distribution
	and this is the application under consideration here.
	The appropriate choice of $\gamma$ optimizes the efficiency of sampling.
	A measure of this is the effective sample size
	$N/\tau$ where $N$ is the number of samples and $\tau$
	is the integrated autocorrelation time.
	The latter is, however, defined in terms of an observable.
	An observable is an expectation of a specified function
	of the configuration, which for lack of a better term,
	is referred to here as a {\em preobservable}.
	As an added complication,
	the accuracy of an estimate of an integrated autocorrelation time (IAcT)
	depends on sampling thoroughness~\cite[Sec.~3]{FaCS20},
	so a conservative approach is indicated.
	Ref.~\cite[Sec.~3.1]{FaCS20} advocates the use of the maximum possible IAcT
	and shows how it might be a surrogate for sampling thoroughness.
	The maximum possible IAcT is about the same (except for a factor of 2)
	as the {\em decorrelation time} of Ref.~\cite{LyZu07},
	defined to be ``the minimum time that must elapse between configurations
	for them to become fully decorrelated (i.e., with respect to any quantity)''.
	
	\topic{optimal gamma}
	Therefore, for sampling, it is suggested that $\gamma$ be chosen
	to achieve a high level of sampling thoroughness,
	as measured by the maximum possible IAcT.
	An initial study of this question is reported in Ref.~\cite[Sec.~5]{SkFa17},
	and the purpose of the present article is to clarify and extend these results.
	% (Unfortunately, Refs.~\cite{FaCS20,SkFa17} contain errors, which are corrected here in \ref{app:FaCS} and \ref{app:SkFa}.)
	
	To begin with, we analyse an underdamped Langevin equation with a quadratic potential energy function. (See Eq.~(\ref{eq:model}) below.)  
	The main purpose of analyzing this model problem is, of course, to obtain
	insight and heuristics that can be applied to general potential energy functions.
	Needed for choosing the optimal gamma is a substitute for the lowest frequency.
	For the model problem, this can be obtained from the covariance matrix for the
	position coordinates, which is not difficult to compute
	for a general potentials.
	And for estimating  $\tau_{\rmq,\max}$,
	the analysis suggests using the set of all quadratic polynomials,
	which can be achieved using the algorithm of reference~\cite[Sec.~3.5]{FaCS20}.
	
	For molecular simulation, the suggestion is
	that one might choose linear combinations of functions
	of the form $|\vecr_j -\vecr_i|^2$
	and $(\vecr_j -\vecr_i)\cdot(\vecr_k -\vecr_i)$ where each $\vecr_i$
	is an atomic position or center of mass of a group of atoms.
	Such functions share with the potential energy function the property
	of being invariant under a rigid body movement.

	\subsection{Results and discussion}
	
	\topic{model problem analyzed; 2 benefits}
	Section~\ref{sec:model} analyzes integrated autocorrelation times
	for the standard model problem of a quadratic potential energy function.
	An expression is derived for the IAcT for any preobservable;  
	this is applied in Sec.~\ref{ss:tauEst} to check the accuracy of a method for estimating the IAcT.
	In Sec.~\ref{sec:model}, we also determine the maximum IAcT,
	denoted by $\tau_{\rmq,\max}$,
	over all preobservables defined on configurations,
	as well as the damping coefficient $\gamma^\ast$
	that minimizes $\tau_{\rmq,\max}$.
	It is shown that it is polynomials of degree $\le 2$ that produce the
	largest value of $\tau_{\rmq,\max}$.
	And that choosing $\gamma$ equal to the lowest
	frequency, which is half of the optimal value of $\gamma$ for that frequency,
	minimizes $\tau_{\rmq,\max}$.
	These results extend those of Ref.~\cite[Sec.~5]{SkFa17},
	which obtains a (less relevant) result for preobservables
	defined on phase space rather than configuration space.
	
	Sections~\ref{sec:LeMa} and~\ref{sec:threeG} test the heuristics derived
	from the quad\-ratic potential energy on some
	simple potential energy functions giving rise to multimodal
	distributions.
	% on the choice of $\gamma$ that minimizes the max IAcT
	% over all possible preobservables defined on configurations
	Results suggest that the heuristics for choosing the maximizing preobservable
	and optimal gamma are effective.
	
	One of the test problems is one constructed by Ref.~\cite{LeMa13}
	to demonstrate the superiority of BAOAB over other Langevin integrators.
	Experiments for this problem in Sec.~\ref{sec:LeMa}
	are consistent with this claim of superiority.
	
	In defining ``quasi-reliability'' and the notion of thorough sampling,
	Ref.~\cite{FaCS20} makes an unmotivated leap from maximizing over
	preobservables that are indicator functions
	to maximizing over arbitrary preobservables.
	The test problem of Sec.~\ref{sec:threeG}
	provides a cursory look at this question,
	though the matter may warrant further study.
	%This is difficult to justify except in the case of the model problem,
	%for which is might be argued that an indicator function can be arbitrarily
	%well approximated by a linear combination of Hermite polynomials.
	%It is not at all clear that this argument works for other less
	%special Langevin propagators.
	% Is the leap misguided or does it give a good ballpark estimate of the
	% max tau over all indicator functions
	%Following Ref.~\cite{LyZu07}, try the three conformations
	
	Obtaining reliable estimates of the IAcT
	without generating huge sets of samples
	very much hinders this investigation.
	To this end, Sec.~\ref{ss:alt} explores an intriguing way of calculating
	an estimate for the phase space $\tau_{\max}$,
	which avoids the difficult calculation of IAcTs.
	For the model problem, it give more accurate results for $\tau_{\max}$
	than estimating IAcTs, due to the difficulty of finding a set of functions that play the same role
	as quadratic polynomials when maximizing IAcTs.
	The literature offers interesting suggestions that might help
	in the development of better schemes for estimating IAcTs,
	and it may be fruitful to
	recast some of these ideas using the formalisms employed in this article.
	In particular, Ref.~\cite{LyZu07} offers a novel approach based on
	%testing independence of the samples in a set.
	determining whether using every $\tau$\,th
	sample creates a set of independent samples.
	Additionally,
	there are several conditions on covariances~\cite[Theorem~3.1]{Geye92}
	that can be checked or enforced.

	\subsection{Related work}
	
	While the major part of the  literature on Markov chain Monte Carlo (MCMC) methods with stochastic differential equations focuses on the overdamped Langevin equation (e.g. \cite{MCMCgrad2002,Betancourt2018} and the references given there), there have been significant advances, both from an algorithmic and a theoretical point of view, in understanding the underdamped Langevin dynamics  \cite{Robert2018}. For example, in Refs. \cite{uberuaga2004,ceriotti2010} Langevin dynamics has been studied from the perspective of thermostatting and enhancment of specific vibrational modes or correlations, in Refs. \cite{Cheng2018,Gitman2019,adLaLa2019} Langevin dynamics has been used to tackle problems in machine learning and stochastic optimisation. From a theoretical point of view, the Langevin equation is more difficult to analyse than its overdamped counterpart, since the noise term is degenerate and the associated propagator is non-symmetric; recent work on optimising the friction coefficient for sampling is due to \cite{Duncan2017,roussel2018,HighLow2020}, theoretical analyses using both probabilistic and functional analytical methods have been conducted in \cite{Dolbeaut2015,Cao2019,Eberle2019}; see also \cite[Secs.~2.3--2.4]{ANUM2016} and the references therein.

	Relevant in this regard are Refs.~\cite{Hwang2005,Lelievre2013,ReyBellet2015}, in which non-reversible perturbations of the overdamped Langevin equation are proposed, with the aim of increasing the spectral gap of the propagator or reducing the asymptotic variance of the sampler. Related results on decorrelation times for the overdamped Langevin using properties of the dominant spectrum of the infinitesimal generator of the associated Markov process have been proved in \cite[Sec.~4]{lebris2012}.
	
	A key point of this article is that quantities like spectral gaps or asymptotic variances are not easily accessible numerically, therefore computing goal-oriented autocorrelation times (i.e. for specific observables that are of interest) that can be computed from simulation data is a sensible approach. With that being said, it would be a serious omission not to mention the work of Ref.~\cite{LyZu07}, which proposes the use of indicator functions for subsets of configuration space in order to estimate asymptotic variance and effective sample size from autocorrelation times using trajectory data.

	Finally, we should also mention that many stochastic optimisation methods that are nowadays popular in the machine learning comminity, like ADAM or RMSProp, adaptively control the damping coefficient, though in an \emph{ad-hoc} way, so as to improve the convergence to a local minimum. They share many features with adaptive versions of Langevin thermostats that are used in moecular dynamics \cite{Matthews2015}, and therefore it comes as no surprise that the Langevin model is the basis for the stochastic modified equation approach that can be used to analyse state of the art momentum-based stochastic optimisation algorithms like ADAM \cite{An2019,SME2019}.

	\section{Preliminaries}  \label{sec:prelim}
	
	\topic{aim}
	The computational task
	is to sample from a probability density $\rho_\rmq(\bfq)$
	proportional to $\exp(-\beta V(\bfq))$,
	where $V(q)$ is a potential energy function and $\beta$ is inverse temperature.
	In principle, these samples are used to compute an observable
	$\bbE[u(\bfQ)]$, where $\bfQ$ is a random variable from the prescribed
	distribution and $u(\bfq)$ is a preobservable
	(possible an indicator function).
	The standard estimate is
		\[
		\bbE[u(\bfQ)]\approx\widehat{U}_N =\frac1{N}\sum_{n=0}^{N-1} u(\bfQ_n),
		\]
		where the samples $\bfQ_n$ are from a Markov chain,
		for which $\rho_\rmq(\bfq)$ (or a close approximation thereof) is the stationary density.
		Assume the chain has been equilibrated,
		meaning
		that $\bfQ_0$ is drawn from a distribution with density $\rho_\rmq(\bfq)$.
	An efficient and popular way to generate such a Markov chain
	is based on Langevin dynamics,
	whose equations are
	\begin{equation} \label{eq:LD}
		\begin{aligned}
			\rmd\bfQ_t & = M^{-1}\bfP_t\,\rmd t,\\
			\rmd\bfP_t & = F(\bfQ_t)\,\rmd t -\gamma\bfP_t\,\rmd t
			+\textstyle\sqrt{\frac{2\gamma}{\beta}}\Mh\,\rmd\bfW_t,
		\end{aligned}
	\end{equation}
	where $F(\bfq) = -\nabla V(\bfq)$,
	$M$ is a matrix chosen to compress the range of vibrational frequencies,
	$\Mh\Mh\T = M$, and
	$\bfW_t$ is a vector of independent standard Wiener processes.
	The invariant phase space probability density $\rho(\bfq,\bfp)$ is given by 
	\[
	\rho(\bfq,\bfp) = \frac{1}{Z}\exp(-\beta(V(\bfq) +\frac12\bfp\T M^{-1}\bfp)),
	\]
	where $Z>0$ is a normalisation constant that guarantees that $\rho$ integrates to 1. We call $\rho_\rmq(\bfq)$ its marginal density for $\bfq$. We suppose $\rho>0$. 
	
	\topic{example}
	It is common practice in molecular dynamics to use a numerical integrator,
	which introduces a modest bias, that depends on the step size $\Delta t$.
	As an illustration, consider the BAOAB integrator~\cite{LeMa13}.
	Each step of the integrator consists of the following substeps:
	\begin{itemize}
		\item[B:] $\bfP_{n+1/4} =\bfP_n +\frac12\Delta t F(\bfQ_n)$,
		\item[A:] $\bfQ_{n+1/2} =\bfQ_n +\frac12\Delta t M^{-1}\bfP_{n+1/4}$,
		\item[O:] $\bfP_{n+3/4} =\exp(-\gamma\Delta t)\bfP_{n+1/4} +\bfR_{n+1/2}$,
		\item[A:] $\bfQ_{n+1} =\bfQ_{n+1/2} +\frac12\Delta t M^{-1}\bfP_{n+3/4}$,
		\item[B:] $\bfP_{n+1} =\bfP_{n+3/4} +\frac12\Delta t F(\bfQ_{n+1/2})$,
	\end{itemize}
	where $\bfR_{n+1/2}$
	is a vector of independent Gaussian random variables with mean $\bfzero$
	and covariance matrix $(1 -\exp(-2\gamma\Delta t))\beta^{-1}M$.
	
	\topic{variance}
	%The variance of the estimated mean is given by
	%\[
	%\Var[\widehat{U}] =\frac1{N}\Var[u(\bfQ)]
	%\left(1 + 2\sum_{k=1}^{N-1}\left(1 -\frac{k}{N}\right)\frac{C(k)}{C(0)}\right)
	%\]
	%More concisely, 
	In the following, we use the shorthand $\bfZ=(\bfQ,\bfP)$ to denote a phase space vector. It is known~\cite[Sec.~2]{Geye92} that
	the variance of the estimate $\widehat{U}_N$ for $\bbE[u(\bfZ)]$ is
	\begin{equation}  \label{eq:Varbar}
		\Var[\widehat{U}_N]\approx\frac{\tau}{N}\Var[u(\bfZ)],
	\end{equation}
	which is exact relative to $1/N$ in the limit $N\rightarrow\infty$. 
	Here $\tau$ is the {\it integrated autocorrelation time (IAcT)}
	\begin{equation}  \label{eq:tau}
		\tau = 1 + 2\sum_{k=1}^{+\infty}\frac{C(k)}{C(0)}
	\end{equation}
	and $C(k)$ is the autocovariance at lag $k$ defined by
	\begin{equation}  \label{eq:defCk}
		C(k) =\bbE[(u(\bfZ_0) -\mu)(u(\bfZ_k) -\mu)]
	\end{equation}
	with $\mu =\bbE[u(\bfZ_0)] =\bbE[u(\bfZ_k)$. Here and in what follows the expectation $\bbE[\cdot]$ is understood over all realisations of the (discretized) Langevin dynamics, with initial conditions $\bfZ_0$ drawn from the equilibrium probability density function $\rho$.
	
	\subsection{Estimating integrated autocorrelation time} 
	
	Estimates of the IAcT based on estimating covariances $C(k)$
	suffer from inaccuracy in estimates of $C(k)$
	due to a decreasing number of samples as $k$ increases.
	To get reliable estimates, it is necessary to underweight or omit
	estimates of $C(k)$ for larger values of $k$.
	Many ways to do this have been proposed.
	Most attractive are those~\cite[Sec.~3.3]{Geye92} that take advantage
	of the fact that the time series is a Markov chain.
	%As a protection against excessive noise, they employ a windowing function:
	%\[
	%\widehat{\tau}\approx 1 + 2\sum_{k=1}^{N-1}w(k)
	%\frac{\widehat{C}(k)}{\widehat{C}(0)}
	%\]
	%where $w(k)$ is a decreasing weight known as a {\em lag window}.
	
	One that is used in this study is a short computer program
	called {\tt acor}~\cite{Good09} that implements a method described 
	in Ref.~\cite{MaSo88}.
	It recursively reduces the series to one half its length by summing successive pairs of terms
	until the estimate of $\tau$ based on the reduced series is deemed reliable.
	The definition of ``reliable'' depends on heuristically chosen parameters.
	A greater number of reductions, called $\mathit{reducs}$ in this paper, employs
	greater numbers of covariances, but at the risk of introducing more noise.
	
	\subsection{Helpful formalisms for analyzing MCMC convergence}
	
	It is helpful to introduce the linear operator $\calT$ defined by
	\[\calT u(\bfz) =\int\rho(\bfz'|\bfz)u(\bfz')\rmd\bfz'\]
	where $\rho(\bfz'|\bfz)$
	is the transition probability density for the  Markov chain.
	Then one can express an expectation of the form $\bbE[v(\bfZ_0)u(\bfZ_1)]$,
	arising from a covariance, as
	\[\bbE[v(\bfZ_0)u(\bfZ_1)] =\langle v,\calT u\rangle\]
	where the inner product $\langle\cdot,\cdot\rangle$ is defined by
	\begin{equation}  \label{eq:inrprod}
		\langle v, u\rangle
		= \int v(\bfz)u(\bfz)\rho(\bfz)\,\rmd\bfz.
	\end{equation}
	The adjoint operator
	\[\calT^\dagger v(\bfz) =\frac1{\rho(\bfz)}
	\int\rho(\bfz|\bfz')v(\bfz')\rho(\bfz')\rmd\bfz'\]
	is what Ref.~\cite{ScHu00}
	calls the forward transfer operator, because
	it propagates relative probability densities forward in time.
	On the other hand, Ref.~\cite{LiWK94} calls $\calT^\dagger$
	the backward operator and calls $\calT$ itself the forward operator.
	To avoid confusion, use the term {\em transfer operator} for $\calT$.
	The earlier work~\cite{FaCS20,SkFa17} is in terms
	of the operator $\calT^\dagger$.
	To get an expression for $\bbE[v(\bfZ_0)u(\bfZ_k)]$, write
	\[\bbE[v(\bfZ_0)u(\bfZ_k)]
	=\int\!\!\int v(z)u(z')\rho_k(z'|z)\rho(z)\,\rmd z\rmd z'\]
	where $\rho_k(z'|z)$ is the iterated transition probability density function
	defined recursively by $\rho_1(z'|z) = \rho(z|z')$ and
	\[\rho_k(z'|z) =\int\rho(z'|z'')\rho_{k-1}(z''|z)\rmd z''\,, \quad k = 2, 3,\ldots.\]
	By induction on $k$
	\[\calT^k u(z) = \calT\calT^{k-1}u(z)
	=\int\rho_k(z'|z) u(z')\rmd z',\]
	whence,
	\[
	\bbE[v(\bfZ_0)u(\bfZ_k)] =\langle v, \calT^k u\rangle.
	\]
	
	\subsubsection{Properties of the transfer operator and IAcT}

	It is useful to establish some properties of $\calT$ and the IAcT that will be used throughout the article. In particular, we shall provide a formula for $\tau(u)$ in terms of the transfer operator that will be the starting point for systematic improvements and that will later on allow us to estimate $\tau$ by solving a generalised eigenvalue problem.
	
	Clearly, $\calT\,1 = 1$, and 1 is an eigenvalue of $\calT$.
	Here, where the context requires a function, the symbol 1 denotes
	the constant function that is identically 1.
	Where the context requires an operator, it denotes the identity operator.
	To remove the eigenspace corresponding to the eigenvalue $\lambda=1$ from $\calT$,
	define the orthogonal projection operator
	\[\calE u =\langle 1, u\rangle\,1\]
	and consider instead the operator
	\[\calT_0 =\calT -\calE.\]
	It is assumed
		that the eigenvalues $\lambda$ of $\calT_0$ satisfy $|\lambda| < 1$, in other words, we assume that the underlying Markov chain is ergodic.
	Stationarity of the target density $\rho(\bfz)$ w.r.t. $\rho(\bfz|\bfz')$
	implies that $\calT^\dagger\,1 = 1$ and that $\calT^\dagger\calT\,1 = 1$.
	Therefore, $\calT^\dagger\calT$ is a stochastic kernel.
	This implies that the spectral radius of $\calT^\dagger\calT$ is 1,
	and, since it is a symmetric operator, one has that
	\begin{equation}  \label{eq:normTau}
		\langle\calT u,\calT u\rangle =\langle u,\calT^\dagger\calT u\rangle
		\le\langle u, u\rangle.
	\end{equation}
	
	The IAcT, given by Eq.~(\ref{eq:tau}), requires autocovariances,
	which one can express in terms of $\calT_0$ as follows:
	\begin{equation}  \label{eq:Ck}
		\begin{aligned}
			C(k) =&\langle(1 -\calE) u, (1 -\calE)\calT^k u\rangle\\
			=&\langle(1 -\calE) u, (1 -\calE)\calT_0^k u\rangle \\
			=&\langle(1 -\calE) u,\calT_0^k u\rangle, 
		\end{aligned}
	\end{equation}
	which follows because $\calE$ and $1 -\calE$ are symmetric.
	Substituting Equation~(\ref{eq:Ck}) into Equation~(\ref{eq:tau})
	gives
	\begin{equation}  \label{eq:tauOFu}
		\tau(u) =
		\frac{\langle(1 -\calE)u,\calD u\rangle}%
		{\langle(1 -\calE)u, u\rangle},
		\quad\mbox{where }\;\calD = 2(1 - \calT_0)^{-1} - 1.
	\end{equation}
	It can be readily seen that $\tau$ is indeed nonnegative. With $v = (1 -\calT_0)^{-1}u$,
		the numerator in Eq.~(\ref{eq:tauOFu}) satisfies
		\begin{eqnarray*}\langle(1 -\calE)u,\calD u\rangle
			&=&\langle(1 -\calE)(1 -\calT_0)v, (1 +\calT_0)v\rangle \\
			&=&\langle v, v\rangle -\langle\calT v,\calT v\rangle\\ & \ge & 0.
		\end{eqnarray*}
		Therefore, $\tau(u)\ge0$ if $(1-\calE)u\ne 0$, where the latter is equivalent to $u\neq \bbE[u]$ being not a constant.

	\section{Sampling Thoroughness and Efficiency}
	
	Less than ``thorough'' sampling can degrade estimates of an IAcT.
	Ref.~\cite[Sec.~1]{FaCS20} proposes a notion of ``quasi-reliability''
	to mean the absence of evidence in existing samples
	that would suggest a lack of sampling thoroughness.
	A notion of sampling thoroughness begins by considering
	subsets $A$ of configuration space.
	The probability that $\bfQ\in A$ can be expressed as the expectation
	$\bbE[1_A]$ where $1_A$ is the indicator function for $A$.
	A criterion for thoroughness might be that
	\begin{equation}  \label{eq:tol}
		|\widehat{1_A} -\Pr(\bfQ\in A)|\le\mathit{tol}\quad
		\mbox{where }\widehat{1_A} =\frac1{N}\sum_{n=1}^N 1_A(\bfQ_n).
	\end{equation}
	This is not overly stringent,
	since it does not require that there are any samples
	in sets $A$ of probability $\le\mathit{tol}$.
	
	The next step in the development of this notion
		is to replace the requirement $|\widehat{1_A} -\Pr(\bfQ\in A)|\le\mathit{tol}$
		by something more forgiving of the random error in $\widehat{1_A}$. For example, we could 
		require instead that
		\[(\Var[\widehat{1_A}])^{1/2}\le 0.5 \,\mathit{tol},\]
		which would satisfy Eq.~(\ref{eq:tol}) with 95\% confidence, 
		supposing an approximate normal distribution for the estimate. (If we are not willing to accept the Gaussian assumption, Chebychev's inequality tells us that we reach 95\% confidence level if we replace the right hand side by $0.05\,\mathit{tol}$.)

	Now let $\tau_A$ be the integrated autocorrelation time for $1_A$.
	Because
	\begin{eqnarray*}
		\Var[\widehat{1_A}]
		&\approx&\tau_A\frac1{N}\Var[1_A(\bfZ)] \\
		&=&\tau_A\frac1{N}\Pr(\bfZ\in A)(1 -\Pr(\bfZ\in A))\\
		&\le&\frac1{4N}\tau_A,
	\end{eqnarray*}
	it is enough to have $(1/4N)\tau_A\le(1/4)\mathit{tol}^2$ for all sets
	of configurations $A$ to ensure thorough sampling (assuming again Gaussianity).
	The definition of good coverage might then be expressed in terms
	of the maximum $\tau(1_A)$ over all $A$. 
	Note that the sample variance may not be a good criterion if all the candidate sets $A$  have small probability $\Pr(\bfZ\in A)$, in which case it is rather advisable to consider the \emph{relative} error \cite{tuffin2014}.

	Ref.~\cite[Sec~3.1]{FaCS20} then makes a leap, for the sake of simplicity,
	from considering just indicator functions to arbitrary functions.
	This leads to defining
	$\tau_{\rmq,\max} =\sup_{\Var[u(\bfQ)] > 0}\tau(u)$.
	The condition $\Var[u(\bfQ)] > 0$ is equivalent to $(1 -\calE)u\ne 0$.\\
	
	A few remarks on the efficient choice of preobservables are in order. 
	
	\begin{remark}
		Generally, if there are symmetries present in both the distribution and the preobservables
		of interest, this may reduce the amount of sampling needed.
		Such symmetries can be expressed as bijections $\psi_\rmq$ for which
		$u(\psi_\rmq(\bfq)) = u(\bfq)$ and 
		$\rho_\rmq(\psi_\rmq(\bfq)) =\rho_\rmq(\bfq)$.
		Examples include translational and rotational invariance, as well as
		interchangeability of atoms and groups of atoms.
		Let $\Psi_\rmq$ denote the set of all such symmetries.
		The definition of good coverage then need only include sets $A$,
		which are invariant under all symmetries $\psi_\rmq\in\Psi_\rmq$. The
		extension from indicator sets $1_A$ to general functions leads to considering
		$W_\rmq =\{u(\bfq)~|~
		u(\psi_\rmq(\bfq)) = u(\bfq)$ for all $\psi_\rmq\in\Psi_\rmq\}$
		and defining
			\[\tau_{\rmq,\max} =\sup_{u\in W^0_{\rmq}}\tau(u)\]
			where $W_{\rmq}^0 = \{u\in W_\rmq~|~\Var[u(\bfQ)] > 0\}$.
	\end{remark}

	\begin{remark}
		Another consideration
		that might dramatically reduce the set of relevant preobservables
		is the attractiveness of using collective variables
		$\zeta =\xi(q)$ to characterize structure and dynamics of molecular systems.
		This suggests considering only functions defined on collective variable space,
		hence, functions of the form $\bar{u}(\xi(q))$.
	\end{remark}

	\section{Computing the Maximum IAcT}
	
	The difficulty of getting reliable estimates for $\tau(u)$ in order to compute the maximum IAcT makes it interesting to consider alternative formulation.
	
	\subsection{A transfer operator based formulation}  \label{ss:alt}
	
	Although, there
	is little interest in sampling functions of auxiliary variables like momenta,
	it may be useful to consider phase space sampling efficiency.
	Specifically, a maximum over phase space is an upper bound and it might
	be easier to estimate.
	Putting aside exploitation of symmetries, the suggestion
	is to using $\tau_{\max} =\sup_{\Var[u(\bfZ)] > 0}\tau(u)$.
	One has, with a change of variables, that
	%one can show that
	\[
	\tau((1 - \calT_0)v) =\tau_2(v)
	\]
	where
	\[
	\tau_2(v) =\frac{\langle(1 -\calT)v, (1 +\calT)v\rangle}%
	{\langle(1 -\calT)v, (1 -\calT)v\rangle}.
	\]
	This follows from
	$\langle(1 -\calE)(1 -\calT_0)v, (1\pm\calT_0)v\rangle
	= \langle(1 -\calT)v, (1\pm\calT)v\mp\calE v\rangle
	=\langle(1 -\calT)v, (1\pm\calT)v\rangle$.
	Therefore,
	\begin{eqnarray*}
		\tau_{\max}&=&\sup_{\Var[(1 -\calT_0)v(\bfZ)] > 0}\tau((1 -\calT_0)v) \\
		&=&\sup_{\Var[(1 -\calT_0)v(\bfZ)] > 0}\tau_2(v)\\
		&=&\sup_{\Var[v(\bfZ)] > 0}\tau_2(v).
	\end{eqnarray*}
	The last step follows because $(1 -\calT_0)$ is nonsingular.
	
	\topic{most general case}
	Needed for an estimate of $\tau_2(v)$
	is $\langle\calT v,\calT v\rangle$.
	To evaluate $\langle\calT v,\calT v\rangle$, proceed as follows:
	Let $\bfZ'_{n+1}$ be an independent realization of $\bfZ_{n+1}$ from $\bfZ_n$.
	In particular, repeat the step,
	but with an independent stochastic process having the same distribution.
	Then
		\begin{equation}  \label{eq:EZpZ}
			\begin{aligned}
				\bbE[v(\bfZ_1)v(\bfZ'_1)]
				= & \int\!\int v(z)v(z')\int\rho(z|z'')\rho(z'|z'')\rho(z'')\rmd z''\,\rmd z\rmd z' \\
				= & \langle\calT v,\calT v\rangle. 
			\end{aligned}
	\end{equation}
	For certain simple preobservables and propagators having the simple form of BAOAB, 
		the samples $v(\bfZ_n)v(\bfZ'_n)$ might be obtained at almost no extra cost,
		and their accuracy improved and their cost reduced
		by computing conditional expectations analytically.
	
	This approach has been tested on the model
		problem of Sec.~\ref{sec:model}, a Gaussian process, and found to be significantly better
		than the use of {\tt acor}. Unfortunately, this observation is not generalisable: For example, for a  double well potential, it is difficult to find preobservables $v(\bfz)$, giving a computable estimate of $\tau_{\max}$ which comes close to an estimate from using {\tt acor} with $u(\bfz) = z_1$.

	Another drawback is that the estimates, though computationally inexpensive, require accessing
	intermediate values in the calculation of a time step,
	which are not normally an output option of an MD program. Therefore we will discuss alternatives in the next two paragraphs.
	
	\subsection{A generalised eigenvalue problem}\label{ssec:evp}
	
	Let $\bfu(\bfz)$ be a row vector of arbitary basis functions $u_i(\bfz)$, 
		$i = 1,2,\ldots,\mathit{imax}$ that span a closed subspace of the Hilbert space associated with the inner product $\langle\cdot,\cdot\rangle$ defined by (\ref{eq:inrprod}) 
		and consider the linear combination $u(\bfz) =\bfu(\bfz)\T\bfx$.
	One has
	\[\tau(u)
	=\frac{\langle(1 -\calE)u,\calD u\rangle}%
	{\langle(1 -\calE)u, u\rangle}
	=\frac{\bfx\T\bfD\bfx}{\bfx\T\bfC_0\bfx}\]
	where
	\[\bfD = \langle(1 -\calE)\bfu,\calD \bfu\T\rangle
	\quad\mbox{and}\quad\bfC_0 =\langle(1 -\calE)\bfu,\bfu\T\rangle.\]
	If the span of the basis is sufficiently extensive to include preobservables
		having the greatest IAcTs (e.g. polynomials, radial basis functions, spherical harmonics, etc.), the calculation of $\tau_{\max}$ reduces to that
		of maximizing $\bfx\T\bfD\bfx/(\bfx\T\bfC_0\bfx)$ over all $\bfx$,
		which is equivalent to solving the symmetric generalized eigenvalue problem
		\begin{equation}\label{eq:evp}
			\frac12(\bfD +\bfD\T)\bfx = \lambda\bfC_0\bfx.
	\end{equation}

	It should be noted that the maximum over all linear combinations
	of the elements of $\bfu(\bfz)$
	can be arbitrarily greater than use of any of the basis functions
	individually. Moreover, in practice, the coefficients in (\ref{eq:evp}) will be random in that they have to be estimated from simulation data, which warrants special numerical techniques. These techniques, including classical variance reduction methods, Markov State Models or specialised basis functions, are not the main focus of this article and we therefore refer to the articles   \cite{Zuckerman2009,nueske2013}, and the references given there.
	
	\begin{remark}
		\ref{app:sym} records different notions of reversibility of the transfer operator that entail specific restrictions on the admissible basis functions that guarantee that the covariance matrices, and thus  $\bfC_0$, remain symmetric.	
	\end{remark}
	
	\subsection{The use of {\tt acor}}\label{ssec:acor}

	It is not obvious how to use an IAcT estimator to construct matrix off-diagonal
	elements $D_{ij} =\langle(1 -\calE)u_i,\calD u_j\T\rangle$, $j\ne i$,
	from the time series $\{\bfu(\bfZ_m)\}$. Nevertheless, it makes sense to use {\tt arcor} as a preprocessing or predictor step to generate an initial guess for an IAcT. The {\tt acor} estimate for a scalar preobservable $u(\bfz)$ has the form
		\[\widehat{\tau} = \widehat{D}/\widehat{C}_0\]
		where
		\[\widehat{C}_0 =\widehat{C}_0(\{u(\bfZ_n) -\hat{U}\},\{u(\bfZ_n) -\hat{U}\})\]
		and
		\[\widehat{D} =\widehat{D}(\{u(\bfZ_n) -\hat{U}\},\{u(\bfZ_n) -\hat{U}\})\]
		are bilinear functions of their arguments that depend
		on the number of reductions $\mathit{reducs}$ where $\hat{U}$ denotes the empirical mean of $\{\bfu(\bfZ_m)\}$.
	
	The tests reported in Secs.~\ref{sec:model}--\ref{sec:threeG} then use the following algorithm. (In what follows we assume that $\{\bfu(\bfZ_m)\}$ has been centred by subtracting the empirical mean.)
	%
	%\begin{figure}[h!]
	\begin{algorithm}[H]
		\caption{Computing the IAcT}\label{reducs1}
		\begin{spacing}{1.1}
			\begin{algorithmic}
				\State For each basis function, compute $\widehat{\tau}$, and record the number of reductions, set $\mathit{reducs}$ to the maximum of these.
				
				\State Then compute $\bfD=(D_{ij})_{ij}$ from $\widehat{D}(\{u_i(\bfz_m)\},\{u_j(\bfz_n)\})$
				with a number of reductions equal to $\mathit{reducs}$.
				
				\If {$\bfD +\bfD\T$ has a non-positive eigenvalue} 
				\State 
				redo the calculation using $\mathit{reducs}-1$ reductions.
				\EndIf
			\end{algorithmic}
		\end{spacing}
	\end{algorithm}
	%\end{figure}

	Ref.~\cite[Sec.~3.5]{FaCS20} uses a slightly different
	algorithm that proceeds as follows:
	%
	%\begin{figure}[h!]
	\begin{algorithm}[H]
		\caption{Computing the IAcT as in \cite[Sec.~3.5]{FaCS20}}\label{reducs2}
		\begin{spacing}{1.1}
			\begin{algorithmic}
				\State Set $\mathit{reducs}$ to the value of $\mathit{reducs}$
				for the basis function having the largest estimated IAcT.
				
				\State Then run {\tt acor} with a number of reductions equal to $\mathit{reducs}$
				to determine a revised $\bfD$ and a maximizing $\bfx$.
				
				\State For $\bfu\T\bfx$, determine the number of reductions $\mathit{reducs}'$.
				
				\If {$\mathit{reducs}'<\mathit{reducs}$}, 
				\State redo the calculation with
				$\mathit{reducs}=\mathit{reducs}'$ and repeat until
				the value of $\mathit{reducs}$ no longer decreases.
				\EndIf
			\end{algorithmic}
		\end{spacing}
	\end{algorithm}

	In the experiments reported here, the original algorithm sometimes
	does one reduction fewer than the new algorithm.
	
	\begin{remark}
		Theoretically, the matrix $\bfD +\bfD\T$ is positive definite.
		If it is not, that suggests that the value of $\mathit{reducs}$ is not
		sufficiently conservative, in which case $\mathit{reducs}$ needs to be reduced. A negative eigenvalue might also arise if the Markov chain does not converge due to a stepsize $\Delta t$ that is too large.
		This can be confirmed by seeing whether the negative eigenvalue persists
		for a larger number of samples.
	\end{remark}
	
	\section{Analytical Result for the Model Problem}  \label{sec:model}
	
	The question of optimal choice for the damping coefficient
	is addressed in Ref.~\cite[Sec.~5.]{SkFa17}
	for the standard model problem $F(\bfq) = - K\bfq$,
	where $K$ is symmetric positive definite,
	for which the Langevin equation is
	\begin{equation}   \label{eq:model}
		\begin{aligned}
			\rmd\bfQ_t&= M^{-1}\bfP_t\,\rmd t,\\
			\rmd\bfP_t&= -K\bfQ_t\,\rmd t -\gamma\bfP_t\,\rmd t
			+\textstyle\sqrt{\frac{2\gamma}{\beta}}\Mh\,\rmd\bfW_t. 
		\end{aligned}
	\end{equation}
	Changing variables $\bfQ' =\Mh\T\bfQ$ and $\bfP' =\Mh^{-1}\bfP$
	and dropping the primes gives $\rmd\bfQ_t = \bfP_t\,\rmd t$,
	\[
	\rmd\bfP_t = -\Mh^{-1}K\Mh\iT\bfQ_t\,\rmd t
	-\gamma\bfP_t\,\rmd t
	+\sqrt{2\gamma/\beta}\,\rmd\bfW_t.
	\]
	With an orthogonal change of variables, this decouples into scalar equations,
	each of which has the form
	\[\rmd Q_t = P_t\,\rmd t,\quad
	\rmd P_t
	= -\omega^2 Q_t\,\rmd t -\gamma P_t\,\rmd t +\sqrt{2\gamma/\beta}\,\rmd W_t\]
	where $\omega^2$ is an eigenvalue of $\Mh^{-1}K\Mh\iT$,
	or, equivalently, an eigenvalue of $M^{-1}K$.
	Changing to dimensionless variables $t' =\omega t$, $\gamma' =\gamma/\omega$,
	$Q' = (\beta m)^{1/2}\omega Q$, $P' = (\beta/m)^{1/2}P$,
	and dropping the primes gives
	\begin{equation}  \label{eq:anal}
		\rmd Q_t = P_t\,\rmd t,\quad
		\rmd P_t
		= - Q_t\,\rmd t -\gamma P_t\,\rmd t +\sqrt{2\gamma}\,\rmd W_t.
	\end{equation}
	For an MCMC propagator,
	assume exact integration with step size $\Delta t$.
	From Ref.~\cite[Sec.~5.1]{SkFa17}, one has
	$\calT = (\rme^{\Delta t\calL})^\dagger =\exp(\Delta t\calL^\dagger)$
	where
	\[\calL^\dagger f 
	= p\frac{\partial}{\partial q}f
	- q\frac{\partial}{\partial p}f -\gamma p\frac{\partial}{\partial p}f
	+\gamma\frac{\partial^2}{\partial p^2}f.\]
	The Hilbert space defined by the inner product from Eq.~(\ref{eq:inrprod})
	has, in this case, a decomposition into linear subspaces
	$\bbP_k =\mathrm{span}\{\He_m(q)\He_n(p)~|~m + n = k\}$
	(denoted by $\bbP'_k$ in Ref.~\cite[Sec.~5.3]{SkFa17}).
	Let
	\[\bfu_k\T
	= [\He_k(q)\He_0(p),\,\He_{k-1}(q)\He_1(p),\,\ldots,\,\He_0(q)\He_k(p)],\]
	and, in particular,
	\begin{eqnarray*}
		\bfu_1\T&=&[ q,\,p],\\
		\bfu_2\T&=&[ q^2 - 1,\,qp,\,p^2 - 1],\\
		\bfu_3\T&=&[q^3 - 3q,\,(q^2 - 1)p,\,q(p^2 - 1),\,p^3 - 3p],\\
		\bfu_4\T&=&[q^4 - 6q^2 + 3,
		\,(q^3 - 3q)p,\,(q^2 - 1)(p^2 - 1),\\&&\mbox{}q(p^3 - 3p),\,p^4 - 6p + 3].
	\end{eqnarray*}
	With a change of notation from Ref.~\cite[Sec.~5.3]{SkFa17},
	$\calL\bfu_k\T = \bfu_k\T\bfA_k$, with $\bfA_k$ given by
	\begin{equation}  \label{eq:A}
		\bfA_k = \left[\begin{array}{cccc}
			0 &       1 &        &          \\
			-k & -\gamma & \ddots &          \\
			&  \ddots & \ddots &        k \\
			&         & -1     & -k\gamma \\
		\end{array}\right].
	\end{equation}
	One can show, using arguments similar to those in~\cite[Sec.~5.3]{SkFa17},
	that $\bbP_k$ closed under application of $\calL^\dagger$.
	Therefore, $\calL^\dagger\bfu_k\T =\bfu_k\T\bfB_k$
	for some $k+1$ by $k+1$ matrix $\bfB_k$.
	Forming the inner product of $\bfu_k$ with each side of this equation gives
	$\bfB_k =\bfC_{k,0}^{-1}\langle\bfu_k,\calL^\dagger\bfu_k\T\rangle$
	where
	$\bfC_{k,0}=\langle\bfu_k,\bfu_k\T\rangle$.
	It follows that
	\[\bfB_k =\bfC_{k,0}^{-1}\langle\bfu_k,\calL^\dagger\bfu_k\T\rangle =\bfC_{k,0}^{-1}\langle\calL\bfu_k,\bfu_k\T\rangle\] 
	and
	\[\calL^\dagger\bfu_k\T =\bfu_k\T\bfC_{k,0}^{-1}\bfA_k\T\bfC_{k,0}.\]
	The Hermite polynomials $\bfu_k$ are orthogonal and
	\[\bfC_{k,0} =\mathrm{diag}(k!0!,\,(k-1)!1!,\,\ldots,\,0!k!).\]
	Also, $\calE \bfu_k\T =\bfzero\T$.
	Accordingly,
	\[\calT_0\bfu_k\T =\calT\bfu_k\T =\bfu_k\T\bfC_{k,0}^{-1}\exp(\Delta t\bfA_k\T)\bfC_{k,0}\]
	and
	\begin{equation}  \label{eq:DuT}
		\calD \bfu_k\T =\bfu_k\T\bfC_{k,0}^{-1}\bfD_k
	\end{equation}
	where
	\begin{eqnarray*}
		\bfD_k
		&=&\bfC_{k,0}
		\left(2(I -\bfC_{k,0}^{-1}\exp(\Delta t\bfA_k\T)\bfC_{k,0})^{-1} - I\right) \\
		&=&-\coth(\frac{\Delta t}2\bfA_k\T)\bfC_{k,0}.
	\end{eqnarray*}
	
	A formula for $\tau(u)$ is possible if $u(q)$ can be expanded
	in Hermite polynomials as
	$u =\sum_{k=1}^\infty c_k\He_k$.
	Then, from Eq.~(\ref{eq:DuT}), $\calD\He_k\in\bbP_k$,
	not to mention $\He_k\in\bbP_k$.
	Using these facts and the mutual orthogonality of the subspaces $\bbP_k$,
	it can be shown that
	\begin{equation}  \label{eq:tauSum}
		\tau(u)
		=\frac{\sum_{k=1}^\infty k!c_k^2\tau(\He_k)}{\sum_{k=1}^\infty k!c_k^2}.
	\end{equation}
	From this it follows that $\max_u\tau(u) =\max_k\tau(\He_k)$.
	
	Since $\He_k =\bfu_k\T\bfx$ with $\bfx =[1, 0,\ldots,0]\T$, one has
	\begin{equation}  \label{eq:tauk}
		\tau(\He_k) =(\bfD_k)_{11}/(\bfC_{k,0})_{11}
		= (\coth(-\frac{\Delta t}2\bfA_k))_{11}.
	\end{equation}
	
	Asymptotically
	$\tau(\He_k)
	= -(2/\Delta t)(\bfA_k^{-1})_{11}$,
	in the limit as $\Delta t\rightarrow 0$.
	In particular,
	\begin{equation}  \label{eq:ASone}
		\bfA_1^{-1}= \left[\begin{array}{rrr}
			-\gamma & -1 \\
			1 & 0
		\end{array}\right]
	\end{equation}
	and
	\begin{equation}  \label{eq:AStwo}
		\bfA_2^{-1}= -\frac1{2\gamma}\left[\begin{array}{ccc}
			\gamma^2 + 1 & -2\gamma & 1 \\
			\gamma & 0        & 0 \\
			1 & 0        & 1
		\end{array}\right].
	\end{equation}
	
\begin{figure}[t!]
		\centering 
		\includegraphics[width=0.5\textwidth]{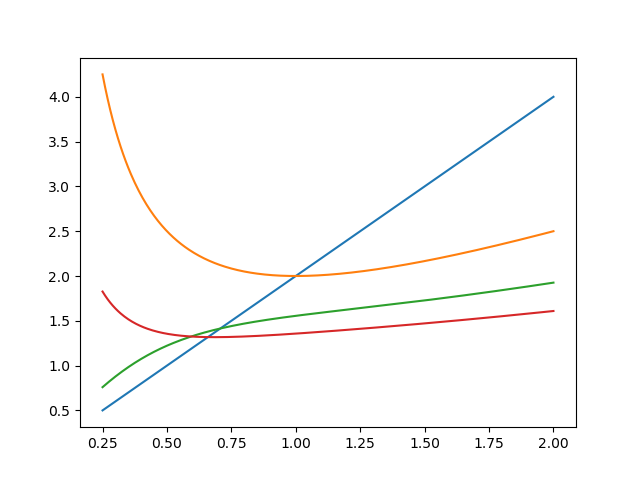}
		\caption{
			From top to bottom on the right $T_k(\gamma)$ vs.\ $\gamma$, $k = 1, 2, 3, 4$
		}
		\label{fig:optxact}
	\end{figure}

	Writing $\tau(\He_k)$ as an expansion in powers of $\Delta t$,
	\[\tau(\He_k)
	= T_k(\gamma)/\Delta t +\calO(\Delta t),\]
	one has $T_1(\gamma) = 2\gamma$ and $T_2(\gamma) =\gamma +1/\gamma$.
	Fig.~\ref{fig:optxact} plots
	$T_k(\gamma)$, $k = 1, 2, 3, 4$, $1/2\le\gamma\le 4$.
	Empirically,
	$\max_k T_k = T_{\max}\stackrel{\mathrm{def}}{=}\max\{T_1, T_2\}$.
	
	Restoring the original variables, one has
	\[\tau_{\rmq,\max}
	= T_{\max}(\gamma/\omega)/(\omega\Delta t) +\calO(\omega\Delta t).\]
	The leading term increases as $\omega$ decreases,
	so $\tau_{\rmq,\max}$
	depends on the lowest frequency $\omega_1$.
	And $\tau_{\rmq,\max}$ is minimized at $\gamma =\omega_1$,
	which is half of the critical value $\gamma = 2\omega_1$.
	Contrast
	this with the result~\cite[Sec.~5.]{SkFa17} for the phase space maximum IAcT,
	which is minimized for $\gamma =(\sqrt{6}/2)\omega_1$.
	
	\begin{remark}
		The result is consistent with related results from \cite{HighLow2020,Eberle2019} that consider optimal damping coefficients that maximise the speed of convergence measured in relative entropy. Specifically, calling $\eta_t=\mathcal{N}(\mu_t,\Sigma_t)$ the law of the solution to (\ref{eq:anal}), with initial conditions $(Q_t,P_t)=(q,p)$; see \ref{app:anal} for details. Then, using \cite[Thm.~4.9]{arnold2014}, we have 
			\[
			KL(\eta_t,\rho) \le M\exp(-2\alpha t)\,,
			\]
			where $M\in(1,\infty)$ and $\alpha$ denotes the spectral abcissa of the matrix $A$ in \ref{app:anal}, 
			i.e. the negative real part of the eigenvalue that is closest to the imaginary axis. Here
			\[
			KL(f,g) = \int\log\frac{f(z)}{g(z)}f(z)\,dz
			\]
			denotes the relative entropy (or: Kullback-Leibler divergence) between two phase space probability densities $f$ and $g$, assuming that 
			\[
			\int_{\{g(z)=0\}} f(z)dz = 0\,.
			\]
			(Otherwise we set $KL(f,g)=\infty$.)
			It is a straightforward calculation to show that the maximum value for $\alpha$ (that gives the fastest decay of $KL(\eta_t,\rho)$) is attained at $\gamma=2$, which is in agreement with the IAcT analysis. For analogous statements on the multidimensional case, we refer to \cite{HighLow2020}.
		
		We should mention that that there may be cases, in which the optimal damping coefficient may lead to a stiff Langevin equation, depending on the eigenvalue spectrum of the Hessian of the potential energy function. As a consequence, optimizing the damping coefficient may reduce the maximum stable step size $\Delta t$ that can be used in numerical simulations.
	\end{remark}

	\subsection{Application to more general distributions}
	
	Note that for the  model problem, the matrix $K$
	can be extracted from
	the covariance matrix
	\[\mathrm{Cov}[\bfQ] = (1/\beta)K^{-1}.\]
	Therefore, as a surrogate for the lowest frequency $\omega_1$,
	and as a recommended value for $\gamma$, consider using
	\[\gamma^\ast = (\lambda_{\min}(M^{-1}K))^{1/2}
	= (\beta\lambda_{\max}(\mathrm{Cov}[\bfQ]M))^{-1/2}\,.\]

	\subsection{Sanity check}  \label{ss:tauEst}
	
	As a test of the accuracy of {\tt acor} and the analytical expression (\ref{eq:tauSum}), the IAcT is calculated by {\tt acor} for a time series generated by the exact analytical propagator (given in \ref{app:anal}) for the reduced model problem given by Eq.~(\ref{eq:model}). For the preobservable, we choose \[
	u(q) =\He_3(q)/\sqrt{3!} -\He_2(q)/\sqrt{2!}
	\]
	where  $\He_2(q) = q^2 - 1$ and $\He_3(q) = q^3 - 3q$ are Hermite polynomials of degree 2 and 3; 
	as damping coefficient, we choose $\gamma = 2$, which is the critical value;
	the time increment is $\Delta t = 0.5$,
	which is about 1/12\,th of a period.
	
\begin{figure}[h!]
		\centering 
		\includegraphics[width=0.5\textwidth]{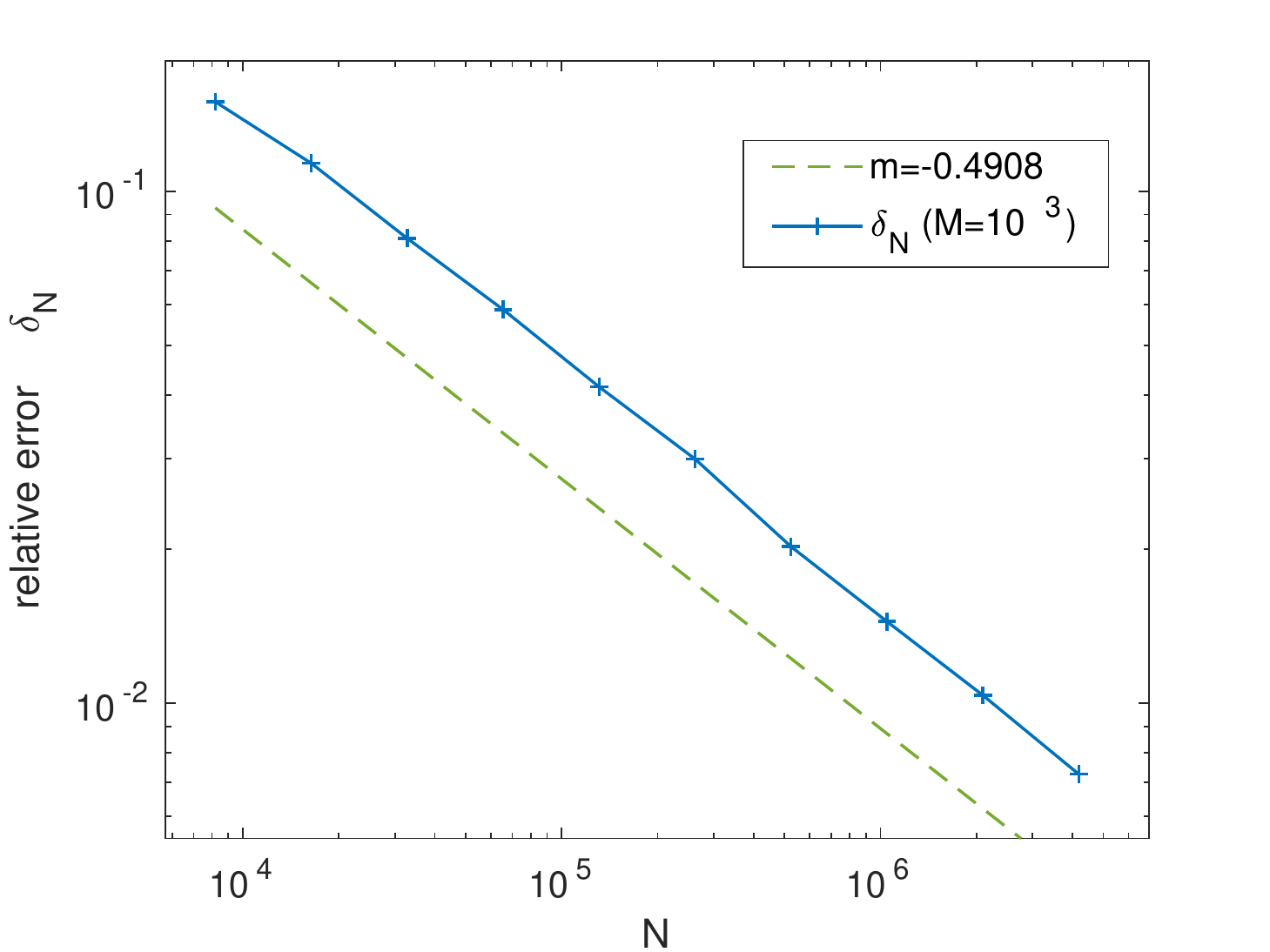}
		\caption{
			Relative error in estimated IAcT $\tau$ as a function of sample size $N$. The relative error $\delta_N  = \sqrt{\Var[\tau]}/\bbE[\tau]$ has been computed by averaging over $M=10^3$ independent realisations of each simulation.}
		\label{fig:loglog}
	\end{figure}

	In this and the other results reported here,
		equilibrated initial values are obtained by running
		for 50\,000 burn-in steps.
		As the dependence of the estimate on $N$ is of interest here, we run $M=10^3$ independent 
		realisations for each value of $N$, from which we can estimate the relative error
		\[
		\delta_N(\tau(u)) = \frac{\sqrt{\Var[\tau(u)]}}{\bbE[\tau(u)]}\,,
		\]
		which we expect to decay as $N^{-1/2}$. Fig.~\ref{fig:loglog} shows the relative error in the estimated IAcT $\tau(u)$ for $N = 2^{13}$, $2^{14}$, \ldots, $2^{22}$. The least-squares fit of the log relative error as a function of $\log N$ has slope $m=0.4908$. Thus we observe a nearly perfect $N^{-1/2}$ decay of the relative error, in accordance with the theoretical prediction.

	\section{A simple example}  \label{sec:LeMa}
	
	The procedure to determine the optimal damping coefficient in the previous section is based on linear Langevin systems. Even though the considerations of Section \ref{sec:model} do not readily generalize to nonlinear systems, it is plausible to use the harmonic approximation as a proxy for more general systems, since large IAcT values are often due to noise-induced metastability, in which case local harmonic approximations inside metastable regions are suitable. 
		For estimating the maximum IAcT, the model problem therefore suggests the
		use of linear, quadratic and cubic functions of the coordinates, where the latter is suitable to capture the possible non-harmonicity of the potential energy wells in the metastable regime.

	The first test problem, which is from Ref.~\cite{LeMa13},
	possesses an asymmetric multimodal distribution.
	It uses $U(q) = \frac14 q^4 +\sin(1+5q)$ and $\beta = 1$,
	and it generates samples using BAOAB with a step size $\Delta t = 0.2$,
	which is representative of step sizes used in Ref.~\cite{LeMa13}.
	Fig.~\ref{fig:eigfuncsLeMa} plots with dotted lines the
	unnormalized probability density function.
	% for $-2\le q\le 2$.
	
	\subsection{Choice of basis}  \label{ss:basLeMa}
	
	A first step is to find a preobservable that produces a large IAcT.
	It would be typical of actual practice
	to try to select a good value for $\gamma$.
	To this end, choose $\gamma =\gamma^\ast = 1.276$,
	%1.276133381635875
	To obtain this value, do a run of sample size $N = 2\cdot 10^6$ using
	$\gamma = 1$, as in one of the tests in Ref.~\cite{LeMa13}.
	
\begin{figure}[t!]
		\centering 
		\includegraphics[width=0.5\textwidth]{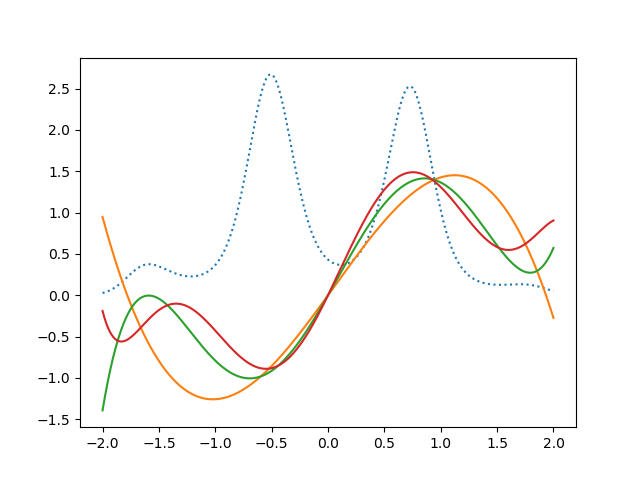}
		\caption{
			In dotted lines is the unnormalized probability density function.
			From top to bottom on the right are the cubic, quintic, and septic
			polynomials that maximize the IAcT over all polynomials of equal degree.
		}
		\label{fig:eigfuncsLeMa}
	\end{figure}

	With a sample size $N = 10^7$, the maximum IAcT is calculated
		for polynomials of increasing degree using the approach described in Secs.~\ref{ssec:evp}--\ref{ssec:acor}.
	Odd degrees produces somewhat greater 
	maxima than even degrees.
	For cubic, quintic, and septic polynomials, $\tau_{\max}$
	has values 59.9, 63.9, 65.8, respectively
	As a check that the sample size is adequate,
	the calculations are redone with half the sample size.
	% 5 runs of 32M steps
	%
	Fig.~\ref{fig:eigfuncsLeMa} shows
	how the maximizing polynomial evolves as its degree increases
	from 3 to 5 to 7.
	
	\subsection{Optimal choice of damping coefficient}
	
	The preceding results indicate that septic polynomials
	are a reasonable set of functions for estimating $\tau_{\rmq,\max}$.
	For 25 values of $\gamma$, ranging from 0.2 to 5, the value
	of $\tau_{\rmq,\max}$ was thus estimated, each run consisting of
	$N = 10^7$ samples.
	
\begin{figure}[t!]
		\centering 
		\includegraphics[width=0.5\textwidth]{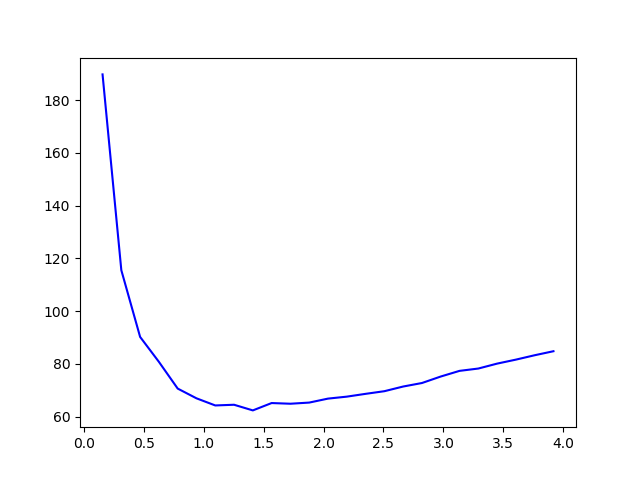}
		\caption{
			$\tau_{\rmq,\max}$ vs.\ $\gamma/\gamma^\ast$ using septic polynomials
			as preobservables   }
		\label{fig:optLeMa}
	\end{figure}

	The optimal value is $\gamma = 1.8 = 1.4\gamma^\ast$,
	which is close the heuristic choice $\gamma^\ast$ for a damping coefficient.
	Fig.~\ref{fig:optLeMa} plots $\tau_{\rmq,\max}$ vs. the ratio $\gamma/\gamma^\ast$.
	
	With respect to this example,
	Ref.~\cite[Sec.~5]{LeMa13} states,
	``We were concerned that the improved accuracy seen in the high $\gamma$ regime
	might come at the price of a slower convergence to equilibrium''.
	The foregoing results indicate
	that the value $\gamma = 1$ used in one of the tests is
	near the apparent optimal value $\gamma = 1.8$.
	Hence, the superior accuracy of BAOAB over other methods
	observed in the low $\gamma$ regime
	does not come at the price of slower convergence.
	
	\section{Sum of three Gaussians}  \label{sec:threeG}
	
\begin{figure}[t!]
		\centering 
		\includegraphics[width=0.5\textwidth]{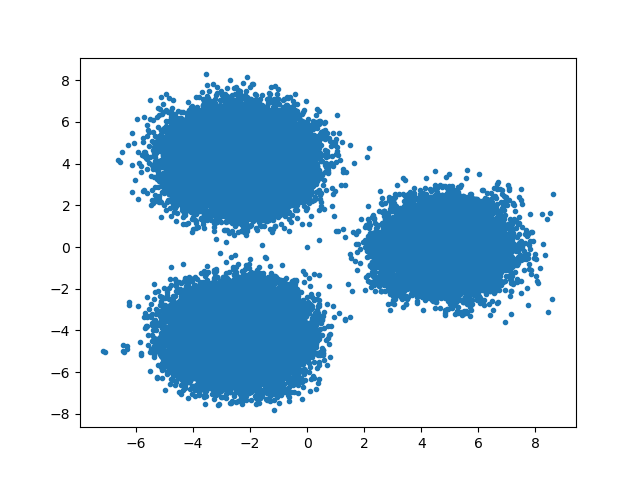}
		\caption{
			A typical time series for a sum of three Gaussians
		}
		\label{fig:trajectory}
	\end{figure}

	The next, perhaps more challenging, test problem uses the sum of
	three (equidistant) Gaussians for the distribution, namely.
	\begin{align*}
		\exp(-V(x, y)) = & \exp(-((x-d)^2 + y^2)/2)\\
		& +\exp(-((x+d/2)^2 + (y-\sqrt{3}d/2)^2)/2) \\
		& +\exp(-((x+d/2)^2 + (y+\sqrt{3}d/2)^2)/2))
	\end{align*}
	where $d$ is a parameter that measures the distance of the three local minima
	from the origin. 
	Integrating the Langevin system using BAOAB with a step size $\Delta t = 0.5$
	as for the model problem, which is what $V(x, y)$ becomes if $d = 0$.
	Shown in Fig.~\ref{fig:trajectory} are the
	first $8\cdot 10^4$ points of a trajectory where $d = 4.8$.
	
	\subsection{Choice of basis}  \label{ss:basthreeG}
	
	To compare $\tau_{\max}$ for different sets of preobservables,
	choose $\gamma =\gamma^\ast = 0.261$,
	and with $\gamma$ so chosen, run the simulation with $d = 4.8$
	for $N = 10^7$ steps.
	To compute $\gamma^\ast$, run the simulation
	for $N = 2\cdot 10^6$ steps
	with $\gamma=1$ (which is optimal for $d = 0$).
	
	Here are the different sets of preobservables and the resulting
	values of $\tau_{\max}$:
	\begin{enumerate}
		\item linear polynomials of $x$ and $y$, for which $\tau_{\max} = 18774$,
		\item quadratic polynomials of $x$ and $y$, for which $\tau_{\max} = 19408$,
		\item linear combinations of indicator functions $\{1_A, 1_B, 1_C\}$ for
		the three conformations
		\begin{align*}
		 A & =\{(x, y)~:~ |y|\le\sqrt{3}x\}\\
		 B &=\{(x, y)~:~ y\ge 0\mbox{ and }y\ge\sqrt{3}x\}\\
		 C &=\{(x, y)~:~ y\le 0\mbox{ and }y\le -\sqrt{3}x\}\,,
		\end{align*}
		for which $\tau_{\max} = 18492$,
		\item $1_A$ alone, for which $\tau = 12087$,
		\item $1_B$ alone, for which $\tau = 5056$,
		\item $1_C$ alone, for which $\tau = 4521$.
	\end{enumerate}
	As consequence of these results, the following section
	uses quadratic polynomials to estimate $\tau_{\rmq,\max}$.
	
	\subsection{Optimal choice of damping coefficient}
	
\begin{figure}[t!]
		\centering 
		\includegraphics[width=0.45\textwidth]{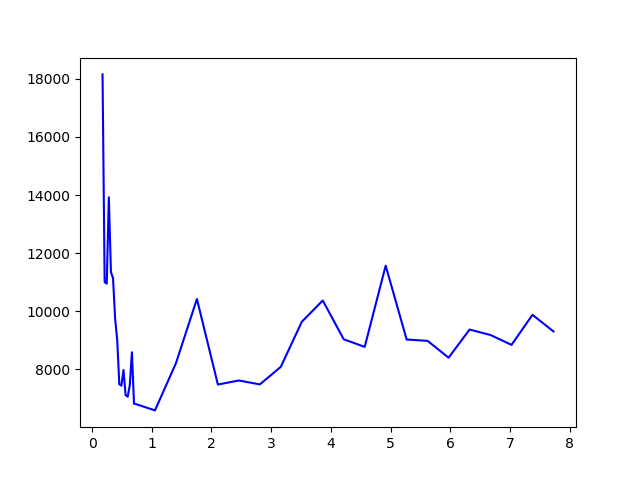}
		\caption{
			$\tau_{\rmq,\max}$ vs.\ the ratio $\gamma/\gamma^\ast$
		}
		\label{fig:optThreegC}
	\end{figure}

	Shown in Fig.~\ref{fig:optThreegC}
		is a plot of $\tau_{\rmq,\max}$ vs.\ the ratio $\gamma/\gamma^\ast$.
		To limit the computing time, we set the parameter to $d = 4.4$
		rather than 4.8 as in Sec.~\ref{ss:basthreeG}; 
		for $d = 4.4$, we have $\gamma^\star = 0.285$,
		obtained using the same protocol as does Sec.~\ref{ss:basthreeG}.
	
	We consider 
		$0.05\le\gamma\le 2.2$ in increments of 0.01 from 0.05 to 0.2, and in increments of 0.1 from 0.2 to 2.2.
		Each data point is based on a run of $N = 2\cdot 10^7$ time steps. Even though the variance of the estimator is not negligible for our choice of simulation parameters, it is clearly visible that the minimum of $\tau_{\rmq,\max}$ is attained at $\gamma\approx \gamma^*$.
	
	\section{Conclusions}
	
	We have discussed the question of how to choose the damping coefficient in (underdamped) Langevin dynamics that leads to efficient sampling of the stationary probability distribution or expectations of certain observables with respect to this distribution. Here, efficient sampling is understood as minimizing the maximum possible (worst-case) integrated autocorrelation time (IAcT).
		We propose a numerical method that is based on the concept of phase space preobservables that span a function space over which the worst-case IAcT is computed using trajectory data; the optimal damping coefficient can then chosen on the basis of this information.
	
	Based on heuristics derived from a linear Langevin equation, we derive rules of thumb for choosing good preobservables for more complicated dynamics. The results for the linear model problem are in agreement with recent theoretical results on Ornstein-Uhlenbeck processes with degenerate noise, and they are shown to be a good starting point for a systematic analysis of nonlinear Langevin samplers.

	\appendix
	
	\section{Analytical propagator for reduced model problem}  \label{app:anal}
	
	This section derives the analytical propagator for Eq.~(\ref{eq:anal}).
	In vector form, the equation is
	\[ \rmd\bfZ_t = A\bfZ\,\rmd t +\bfb\,\rmd W_t\quad\mbox{where }
	A = \left[\begin{array}{cc}
		0 &       1 \\
		-1 & -\gamma
	\end{array}\right]\]
	and $\bfb = [0,\,\sqrt{2\gamma}]\T$.
	The variation of parameters solution is
	\[\bfZ_t =\rme^{t A}\bfZ_0 +\bfR_t\quad\mbox{where }
	\bfR_t =\int_0^t\rme^{(t - s)A}\bfb\,\rmd t.\]
	The stochastic process $\bfR_t$ is Gaussian
	with mean zero and covariance matrix
	\[\Sigma =\bbE[\bfR_t\bfR_t\T]
	=\int_0^t\rme^{(t - s)A}\bfb\bfb\T\rme^{(t - s)A\T}\,\rmd W_t.\]
	To evaluate this expressions, use $A = X\Lambda X^{-1}$ where
	\[
	X = \left[\begin{array}{cc}
		1 &         1 \\
		-\gamma_- & -\gamma_+
	\end{array}\right],
	\quad
	X^{-1} =\frac1{\delta}\left[\begin{array}{ccc}
		\gamma_+ &  1 \\
		-\gamma_- & -1
	\end{array}\right],
	\]
	$\Lambda =\mathrm{diag}(-\gamma_-, -\gamma_+)$,
	\[\gamma_\pm =\frac12(\gamma \pm\delta),\quad\mbox{and}\quad
	\delta =\sqrt{\gamma^2 - 4\omega^2}.\]
	%with the radical sign denoting the principal square root.
	
	Noting that $\exp(-\gamma_\pm t)
	=\exp(-\gamma t/2)(\cosh(\delta t/2)\mp\sinh(\delta t/2))$, one has
	\[\rme^{t A} =\rme^{-\gamma t/2}\cosh\frac{\delta t}2
	\left[\begin{array}{cc} 1 & 0 \\ 0 & 1\end{array}\right]
	+\rme^{-\gamma t/2}\frac{t}2\mathrm{sinhc}\frac{\delta t}2
	\left[\begin{array}{cc} \gamma & 2 \\ -2 & -\gamma\end{array}\right],
	\]
	where $\mathrm{sinhc}\,s =(\sinh s)/s$.
	
	Then
	\begin{eqnarray*}
		\Sigma&=&X
		\int_0^t\rme^{(t - s)\Lambda}X^{-1}\bfb\bfb\T X\iT\rme^{(t - s)\Lambda}\,\rmd t
		X\T \\
		&=&\frac{2\gamma}{\delta^2}X\int_0^t\rme^{(t - s)\Lambda}
		\left[\begin{array}{cc} 1 & -1 \\ -1 & 1\end{array}\right]
		\rme^{(t - s)\Lambda}\,\rmd t X\T \\
		&=&\frac{2\gamma}{\delta^2}X\left[\begin{array}{cc}
			\ds\frac{1 -\rme^{-2\gamma_- t}}{2\gamma_-}
			& \ds -\frac{1 -\rme^{-\gamma t}}\gamma \\
			\ds -\frac{1 -\rme^{\gamma t}}{\gamma}
			& \ds\frac{1 -\rme^{-2\gamma_+ t}}{2\gamma_+}
		\end{array}\right]
		X\T.\end{eqnarray*}
	Noting that $\exp(-2\gamma_\pm t)=\exp(-\gamma t)
	(1 + 2\sinh^2(\delta t/2))\mp2\sinh(\delta t/2)\cosh(\delta t/2))$,
	one has
	\begin{eqnarray*}
		\Sigma&=&(1 -\rme^{-\gamma t})
		\left[\begin{array}{cc} 1 & 0 \\ 0 & 1\end{array}\right]
		-\frac{\gamma t^2}2\rme^{-\gamma t}(\mathrm{sinhc}\frac{\delta t}2)^2
		\left[\begin{array}{cc}\gamma & -2 \\ -2 &\gamma\end{array}\right] \\
		&&\mbox{}
		+\gamma t\rme^{-\gamma t}\mathrm{sinhc}\frac{\delta t}2\cosh\frac{\delta t}2
		\left[\begin{array}{cc} -1 & 0 \\ 0 & 1\end{array}\right].
	\end{eqnarray*}

	%\section{Clarification and corrections for some earlier work} 
	
	\section{Different notions of reversibility}\label{app:sym}
	
	We briefly mention earlier work and discuss different reversiblity concepts for transfer operators. 

	\subsection{Quasi-reversibility}
	
	Ref.~\cite[Sec.~3.4]{FaCS20} introduces a notion of quasi-reversibility.
	A transfer operator $\calT$ is quasi-reversible if
	\[\calT^\dagger = \calR^\dagger\calT\calR\]
	where $\calR$ is an operator such that $\calR^2 = 1$.
	This somewhat generalizes the (suitably modified) definitions
	in Refs.~\cite{FaCS20,SkFa17}.
	The principal example of such an operator is $\calR u = u\circ R$ where
	$R$ is a bijection such that $R\circ R =\mathrm{id}$ and $u\circ R = u$
	for $u\in W$, e.g, momenta flipping.
	
	The value of the notion of quasi-reversibility
	is that it enables the construction of basis
	functions that lead to a matrix of covariances that possesses
	a type of symmetric structure~\cite[Sec.~3.1]{SkFa17}.
	This property is possessed by ``adjusted'' schemes that employ an
	acceptance test, and by the limiting case $\Delta t\rightarrow 0$
	of unadjusted methods like BAOAB.

	\subsection{Modified detailed balance}  \label{app:modBal}

	A quite different generalization of reversibility, termed
	``modified detailed balance'', is proposed in Ref.~\cite{FaSS14} as a tool
	for making it a bit easier to prove stationarity.
	%This is elaborated in the next paragraph.\ref{app:modBal}.

	Modified detailed balance is introduced in Ref.~\cite{FaSS14} as a concept
	to make it easier to prove stationarity.
	In terms of the transfer operator,
	showing stationarity means showing that $\calF\,1 = 1$,
	where $1$ is the constant function 1.
	
	Ref.~\cite[Eq.~(15)]{FaSS14} defines modified detailed balance
	in terms of transition probabilities.
	The definition is equivalent to
	$\calF = \calR^{-1}\calF^\dagger\calR^{-1}$
	under the assumption that $\calR$ preserves the stationary distribution.
	This readily generalizes to
	\begin{equation}  \label{eq:modBal}
		\calF =\calR_2\calF^\dagger\calR_1
	\end{equation}
	where $\calR_1$ and $\calR_2$ are arbitrary except for the assumption 
	that each of them preserve the stationary distribution.
	Stationarity follows from Eq.~(\ref{eq:modBal})
	because $\calF^\dagger\,1 = 1$ for any adjoint transfer operator
	and $\calR_1\,1 =\calR_2\,1 = 1$ by assumption.
	
	Reference~\cite{FaSS14} has errors, which are corrected in Ref.~\cite{FaSS16}.
	%In addition to these, there are the following clarifications:
	%\begin{itemize}
	%\item Sec.~II.D. item (1), say instead
	% ``Let $\mathbf{y}'$ be the result of a random change to $\mathbf{y}$ \ldots.''
	%\item The display after Eq.~(15) would be clearer, had it been written
	% ``Get DISPLAY by replacing $z$ by $R(z)$ and making use of Eq.~(5).''
	%\end{itemize}

	\bibliography{gamma}
	\bibliographystyle{abbrv}
\end{document}